\documentclass{article}

\usepackage{PRIMEarxiv}

\usepackage[utf8]{inputenc} % allow utf-8 input
\usepackage[T1]{fontenc}    % use 8-bit T1 fonts
\usepackage{hyperref}       % hyperlinks
\usepackage{url}            % simple URL typesetting
\usepackage{booktabs}       % professional-quality tables

\usepackage{amsfonts}       % blackboard math symbols
\usepackage{amsmath,amssymb,amsfonts}
\usepackage{algorithmic}
\usepackage{textcomp}
\usepackage[table,xcdraw]{xcolor}

\usepackage{nicefrac}       % compact symbols for 1/2, etc.
\usepackage{microtype}      % microtypography
\usepackage{lipsum}
\usepackage{fancyhdr}       % header
\usepackage{graphicx}       % graphics
\usepackage{subcaption}      % subfigures
\usepackage{cite}
\graphicspath{{media/}}     % organize your images and other figures under media/ folder

\title{Applied Machine Learning to Anomaly Detection in Enterprise Purchase Processes}

\author{ A. Herreros-Martínez\\
	Veolia Group\\
	Alicante, Spain  \\
	\texttt{antonio.herreros@veolia.com} 
	\And
	R. Magdalena-Benedicto, J. Vila-Francés, A.J. Serrano-López \\
	Universidad de Valencia, Research Group IDAL\\
	Valencia, Spain \\
	\texttt{\{rafael.magdalena,joan.vila,antonio.j.serrano\}@uv.es} 
        \And
	S. Pérez-Díaz \\
	University of Alcalá de Henares, Research Group ASYNACS,\\
    Madrid, Spain \\
	\texttt{sonia.perez@uah.es} 
}

\begin{document}

\maketitle

\begin{abstract}
In a context of a continuous digitalisation of processes, organisations must deal with the challenge of detecting anomalies that can reveal suspicious activities upon an increasing volume of data. To pursue this goal, audit engagements are carried out regularly, and internal auditors and purchase specialists are constantly looking for new methods to automate these processes. This work proposes a methodology to prioritise the investigation of the cases detected in two large purchase datasets from real data. The goal is to contribute to the effectiveness of the companies’ control efforts and to increase the performance of carrying out such tasks. A comprehensive Exploratory Data Analysis is carried out before using unsupervised Machine Learning techniques addressed to detect anomalies. A univariate approach has been applied through the z-Score index and the DBSCAN algorithm, while a multivariate analysis is implemented with the k-Means and Isolation Forest algorithms, and the Silhouette index, resulting in each method having a transaction candidates’ proposal to be reviewed. An ensemble prioritisation of the candidates is provided jointly with a proposal of explicability methods (LIME, Shapley, SHAP) to help the company specialists in their understanding.
\end{abstract}
\section{Introduction}
\label{sec:introduction}

The Internal Audit department of a company (normally multinationals groups and/or big-sized entities) is aimed to ensure the correctness and effectiveness of the entities’ processes, its compliance to the approved internal policies and to reduce risks in any form that could be presented \cite{b1}. In order to achieve this goal, the companies’ internal teams conduct audits through on a regular basis defined audit engagements. During their missions, the auditors identify, evaluate and document adequate information to achieve the objectives of the engagement \cite{b2}, carrying out interviews with the auditees and performing a rigorous tracking of evidences supporting the audit findings. 

Currently, auditing still mainly relies on sampling the information (registers, transactions, etc.) to assess the processes’ compliance during the audit engagements \cite{b4}. Consequently, the so-called sampling-risk makes that relevant information in the registers/transactions could remain out of the sampling selection to be reviewed. Additionally, with the growing amount of data, this traditional approach becomes obsolete, and the sampling risk is aggravated \cite{b5}.

Among the business processes, a special interest resides in searching for anomalies or misbehaviours on purchases. Internal audit and purchase managers need to prospect, evaluate, and select the methodologies and IT tools capable of monitoring expenses and discovering relevant information that can highlight an out-of-policy act or, even, fraud \cite{b7,b8}. The goal is to automate processes within the company that help to prioritize the investigation activities according to the level of suspicion of any fact.

In this business context, data analytics has been revealed as a key element in supporting organisations in the challenge of processing and controlling huge quantities of information. Among the various algorithms available, those related to machine learning are known to offer good results in finding relevant insights over structured and unstructured data. Machine learning and data analytics are changing the auditing approaches \cite{b10} and are becoming an important part of the control toolbox for companies.

A key drawback to full population tests, based on hand-crafted rules, is that they will likely be able to find only errors, mistakes or deviations which were already expected or anticipated. Machine learning approaches have the ability to go beyond the results obtained through hand-crafted rules. They can be used to find anomalies in large amounts of data \cite{b10} allowing a later identification of errors, mistakes, circumvented processes or even fraud. Most of the machine learning approaches that have been tested in auditing are based on supervised learning \cite{b14,b15}. In this technique, an algorithm learns to map input data to known targets (labelled data). 
In external auditing, due to reporting regulations, data shares a similar structure across multiple companies (e.g., financial statements). This fact makes easier to obtain data labels for training supervised machine learning models. Regarding internal auditing, on the contrary, labels for processes flagging fraudulent/non-fraudulent cases are not available most of the time, given the huge amount of data or the impossibility of previously labelling the suspicious or fraudulent cases. In essence, traditional internal audit planning uses a risk-based approach, selecting areas, departments, or processes to be audited according to criteria such as the turnover in a business area or how long it has not been audited. The unsupervised algorithms offer a promising choice in this context. Within this algorithm’s category, and among others, clustering techniques have been widely studied in anomaly detection \cite{b18,b19}.

The goals of the present work are: 
\begin{enumerate}
\item[i] investigate and propose a methodology able to detect anomalies in an actual unlabelled purchase dataset, \item[ii] prioritise the cases to be analysed by the organisation’s specialists, and \item[iii] provide them an adequate explicability technique of such cases to support their investigation.\end{enumerate}

\section{Purchase Business Process}

Most businesses rely on other businesses for products and services that help them with their operations. The purchasing process enables an organization to evaluate these business-to-business transactions for efficiency and track their spending. It must be distinguished from the procurement process. Procurement focuses on strategies, such as negotiations and researching while purchasing focuses on the actual act of buying products and services. Despite the control measures, an inadequate purchase process management can overlook more deliberate harmful practices (i.e., purchase fraud) and causing enterprise losses \cite{b22}). An effective purchasing process can help prevent theft, fraud, or irregular spending since it requires documenting all business transactions.

\section{Anomalies detection}

Anomaly detection is the process of finding outliers in a given dataset. Outliers are the data objects that stand out amongst other data objects and do not conform to the expected behaviour in a dataset. Anomaly detection algorithms have broad applications in business, scientific, and security domains where isolating and acting on the results of outlier detection is critical \cite{b23}. For the identification of anomalies, there are many different algorithms, such as classification, regression, and clustering. In case of having elements with known anomalous behaviours, then supervised algorithms can be used for anomaly detection. In addition to supervised algorithms, there are specialized (unsupervised) algorithms whose whole purpose is to detect outliers without any previous knowledge about the dataset \cite{b9}.

Anomalies can be classified into three categories \cite{b24}:

\begin{itemize}
\item \underline{Point Anomalies.} Whether an individual data instance can be considered as anomalous with respect to the rest of the data. 
\item \underline{Contextual Anomalies.} Whether a data instance is anomalous in a specific context, but not otherwise. For instance, within a cluster, the points more distant from the centroid can be considered anomalies within the cluster but not in the entire dataset. 
\item \underline{Collective Anomalies.} Whether a collection of related data instances is anomalous with respect to the entire data set or not. For instance, all the elements belonging to a group in a classification problem.
\end{itemize}

\subsection{Anomaly Detection Techniques (ADTs)}
\subsubsection{Outlier Detection Using Statistical Methods}

Outliers in the data can be identified by creating a statistical distribution model of the data and identifying the data points that do not fit into the model or data points that occupy the ends of the distribution tails \cite{b25}. 
Outliers can be detected based on where the data points fall in the standard normal distribution curve. A threshold for classifying an outlier is specified. Other statistical techniques take multiple dimensions into account, computing other distance metrics (multivariate anomaly). 

\subsubsection{Outlier Detection Using Data Science}

Normally outliers show a specific set of characteristics that can be exploited to locate them. According to the unique set of characteristics that identify an outlier, the detection techniques are:

\begin{itemize}
\item \underline{Distance-based:} By nature, outliers are distant from other data points. If the average distance of the nearest N neighbours is measured, the outliers will have a higher value than other normal data points. Distance-based algorithms utilize this property to identify outliers in the data.
\item \underline{Density-based:} The density of a data point in a neighbourhood is inversely related to the distance to its neighbours. Consequently, outliers are located in low-density areas, while the regular data points are concentrated in high-density areas. 
\item \underline{Distribution-based:} Outliers are the data points that have a low probability of occurrence, and they occupy the tail ends of the distribution curve. 
\item \underline{Clustering:} Outliers, by definition, are not similar to normal data points in a dataset. The outliers can be the lone data points that are not clustered or the outliers forming a far and low-populated cluster.
\item \underline{Classification techniques:} Whether labelled data is previously available, a classification model can be built based on a training (labelled) dataset and tested through the prediction of an unknown dataset (not labelled). The challenge in using a classification model is the availability of previously labelled data. 
\end{itemize}

Each of these techniques has multiple parameters and, hence, a data point labelled as an outlier in one algorithm may not be an outlier in another one. Thus, it is a good practice to rely on multiple algorithms before labelling the outliers. This approach is applied in the present work which, considering the problem statement and the unlabelled input data, will be focused on clustering (k-Means, DBSCAN) and graph techniques (Isolation Forest). 

\subsection{Algorithms}
\subsubsection{k-Means}

k-Means clustering is a distance-based clustering method where the dataset is divided into k number of clusters. It is one of the simplest and most commonly used clustering algorithms. In this technique, the user specifies the number of clusters (k) that the dataset need to be grouped in.
The objective of k-Means clustering is to find a prototype data point (centroid) for each cluster; all the data points are then assigned to the nearest prototype, which then forms a cluster. The centre of the cluster can be the mean of all data objects in the cluster, as in k-Means, or the most represented data object, as in k-medoid clustering \cite{b29}.
The evaluation of k-Means clustering has been done using two methods: i) computing the total of the Sum Squared Errors (SSE), where good models will have low SSE within the cluster and low overall SSE among all clusters. SSE can also be referred to as the average within-cluster distance and can be calculated for each cluster and then averaged for all the clusters. ii) the Silhouette coefficient, proposed by Peter Rousseeuw in 1987 \cite{b35}. The Silhouette value is a measure of how similar an object is to its own cluster (cohesion) compared to other clusters (separation).
The Silhouette ranges from -1 to +1, where a high value indicates that the object is well matched to its own cluster and poorly matched to neighbouring clusters. If most objects have a high value, then the clustering configuration is appropriate. If many points have a low or negative value, then the clustering configuration may have too many or too few clusters.

\subsubsection{Isolation Forest}

The Isolation Forest \cite{b41} algorithm takes advantage of two quantitative properties of anomalies: (i) they are the minority consisting of few instances, and (ii) they have attribute-values that are very different from those of normal instances. These characteristics make them more susceptible to being separated (isolated) from normal instances. 
The isolation process is based on a binary tree structure called the isolation tree (iTree). Because of the susceptibility to isolation, anomalies are more likely to be isolated closer to the root of an iTree, whereas normal points are more likely to be isolated at the deeper end of an iTree. The Isolation Forest method (iForest) builds an ensemble of iTrees for a given dataset; anomalies are those instances which have short average path lengths on the iTrees \cite{b41}. Furthermore, the algorithm has a linear time complexity with a low constant and a low memory requirement, which works well with high-volume data such as in the case of the present study \cite{b24}. 

\section{Materials and Methods}

The present work has been carried out using the KNIME Analytics Platform. The tool KNIME v4.5.2- build March 2022 (Konstanz Information Miner), is a free and open-source data analytics platform. It allows an easy implementation of data process and repeatability in case of feedback improvements. KNIME integrates a wide range of components for machine learning and data mining through its modular data pipelining "Building Blocks of Analytics" concept \cite{b33}. A graphical user interface and use of JDBC allows assembly of nodes blending different data sources, including pre-processing (ETL: Extraction, Transformation, Loading), for modelling, data analysis and visualization without, or with only minimal, programming.

\subsection{Data Acquisition \& Pre-processing}

The data used in the experiment was provided by two companies of a multinational Group and came from real procurement transactions occurred during the year 2021.
Since this data was related to real transactions, to avoid any secrecy problems, categorical data that had an associated numerical identification code were excluded from the dataset. As the file had the identification codes for all of them, this deletion did not mean any information loss. Finally, personal information concerning the requester, the buyer and the approvers of each transaction was received anonymized in order to respect the personal data protection directives of the European GDPR regulation.
The initial source of data was composed globally of 65.712 records with 17 columns, split into two datasets, one per company. Table \ref{tab02} contains the extended descriptive figures, which further reveals the different statistical nature of the transactions, inherent in the internal process organisations within each entity. For example, the number of requesters placing orders is significantly higher in Company1. On the contrary, the number of buyers is higher in Company2. This implies a quite different average of purchase orders and items per requester and buyer.

\begin{table}
\begin{center}
\caption{Source Dataset Description}
\label{table}
\setlength{\tabcolsep}{3pt}
\begin{tabular}{|p{200pt}|p{60pt}|p{60pt}|}

\hline
Figure/ Company                               & Company1     & Company2   \\
\hline
\#Records                                    & 27779        & 38162      \\
\#Purchase Orders 					(OrderID)             & 6898         & 15455      \\
\#Items (ItemID)                             & 25961        & 36300      \\
\#Group Categories                           & 3            & 12         \\
\#Material Categories                        & 23           & 26         \\
\#Unique Item 					descriptions              & 17198        & 12018      \\
\#Vendors                                    & 988          & 1126       \\
\#Requesters                                 & 122          & 63         \\
\#Buyers                                     & 11           & 51         \\
\#Approvers                                  & 76           & 101        \\
Min/Max/Mean \#OrderID 					/ Requester    & 1/56/870     & 1/245/1492 \\
Min/Max/Mean \#ItemID 					/ Requester     & 1/212/11884  & 1/576/4882 \\
Min/Max/Mean 					\#VendorCode / Requester & 1/14/102     & 1/31/137   \\
Min/Max/Mean \#OrderID 					/ Buyer        & 1/627/2141   & 1/303/1492 \\
Min/Max/Mean \#ItemID 					/ Buyer         & 1/2360/11907 & 1/712/4883 \\
Min/Max/Mean 					\#VendorCode / Buyer     & 1/119/382    & 1/38/158   \\
Min/Max/Mean \#OrderID 					/ Approver     & 1/95/819     & 1/153/835  \\
Min/Max/Mean \#ItemID 					/ Approver      & 1/357/11574  & 1/359/3531 \\
Min/Max/Mean 					\#VendorCode per Approver  & 1/25/107     & 1/28/174   \\
Total Purchase Amount                        & 98,9MEUR     & 243,9MEUR  \\
\hline
\end{tabular}
\label{tab02}
    
\end{center}
\end{table}

The preparation of the data (cleaning and transformation) included several tasks aimed to make it ready to the modelling phase. 

The statistical profiling performed to understand the nature of the data disclosed the high number of categorical features, which represents a limitation to applying well-known and robust clustering algorithms. To solve this, different Data Coding techniques were used to transform categorical information into numeric.The categorical data encoding used was the \textbf{Target Encoding} technique \cite{b62}, consisting in substituting each group in a categorical feature with the average response in another target and meaningful variable of the dataset (the order amount in our case). This represents a powerful solution because it avoids generating a high number of features, as is the case of One-Shot Encoding, keeping the dimensionality of the dataset as the original one.The idea behind was to analyse the behaviour of these four strategies of target encoding with the algorithms and compare their results.

Other aspects revealed in these pre-analyses were: i) numerical attributes do not follow a normal distribution, ii) the amount of a purchase order is the most relevant numeric information in the dataset, so it would be used to characterise categorical attributes, iii) the few rows removed due to the presence of missing values, iv) no redundant information was found through correlation measures so feature selection to reduce the problem dimensionality was not possible. 

Then, for every dataset it has been applied the different anomaly detection methods, testing the several options for data grouping (median, avergage, count...) in order to perform feature engineeerin with data, both in univariate and multivariate approach.. This process will point out the more disciminant way and method to detect the anomalies

\section{Results}
\subsection{Univariate Anomaly Detection}

A first approximation of this work has been using a univariate approach where four techniques were selected and applied individually upon the 17 columns:
\begin{itemize}
\item Numeric outliers: those transactions whose column value is outside of the IQR rank (Lower bond: (Q1-1.5IQR) – Upper bond: (Q3+1.5IQR)) has been flagged as an outlier.
\item z-Score: those transactions whose z-Score is greater than 2,5 has been flagged as an outlier. 
\item DBSCAN: those transactions that do not have at least three neighbours within a distance of the unity (Core Points) nor are within a distance of the unity from a Core Point (but have less than 3 neighbours), have been flagged as an outlier.
\item Isolation Forest:  (KNIME node executing the implementation in the sklearn.ensemble Python package) \cite{b42}.
\end{itemize}
     
IQR and IForest techniques performed excellently, but they were discarded as they disclosed a high number of outliers (the adopted criteria were to provide a manageable set of transactions to be reviewed by the company specialists). Furthermore, z-Score and DBSCAN are based on different approaches (distance to mean value and density, respectively) and, according to the results, can provide a complementary point of view. As a result, the transactions (rows) detected as outlier by z-Score or DBSCAN and for, at least, one attribute, have been flagged as univariate outlier. Applying this rule, the resulting dataset for each encoding has a different size (number of transactions -rows-), as shown in Table \ref{tab03}.

\begin{table}
\begin{center}  
\caption{Number of transactions per dataset flagged by z-score or DBSCAN as an univariate outlier vs normal transactions.}
\label{table}
\setlength{\tabcolsep}{3pt}
\begin{tabular}{lllll}

\hline
Company   & Encoding & \#UnivariateOutliers & \#NormalElements & \#TotalDataset \\
\hline
Company1 & Count    & 2286                 & 25435            & 27721          \\
Company1 & Mean     & 4546                 & 23175            & 27721          \\
Company1 & Median   & 2870                 & 24851            & 27721          \\
Company1 & Mode     & 4721                 & 23000            & 27721          \\
Company2 & Count    & 2889                 & 35049            & 37938          \\
Company2 & Mean     & 3656                 & 34282            & 37938          \\
Company2 & Median   & 3431                 & 34507            & 37938          \\
Company2 & Mode     & 3741                 & 34197            & 37938         \\
\hline
\end{tabular}
\label{tab03}
\end{center} 
\end{table}

\subsection{Multivariate Anomaly Detection}

An iterative methodology has been followed, starting with the application to the datasets a well-known and proven algorithm and, in successive steps, new methods and algorithms have been added, enriching the final result. Furthermore, a clustering quality optimization and a preliminary model interpretability testing have been carried out.

\subsubsection{k-Means – Optimised - With univariate outliers}

After performing the pre-processing data tasks, the k-Means model has been applied to the dataset. The univariate outliers have been kept in the dataset as our goal was to detect anomalies, and the univariate flagged transactions could represent a specific cluster. The clustering algorithm uses the Euclidean distance on the selected attributes. As the data is not normalized by the node, the "Normalizer" node is used as a pre-processing step to compute a Gaussian normalization (Z-score).

\begin{figure}[!t]
\centerline{\includegraphics[width=0.5\columnwidth]{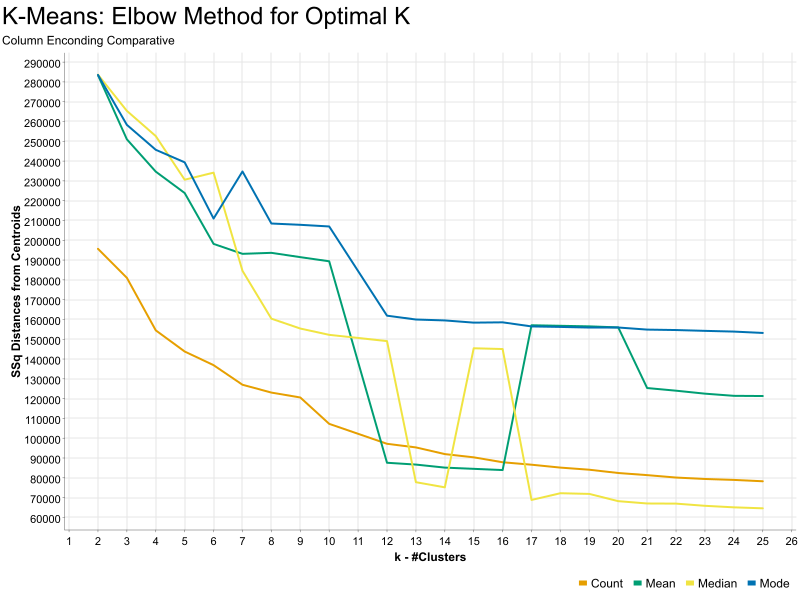}}
\caption{Comparison of the Elbow Curve by encoding strategy (frequency, mean, median and mode). Company 1}
\label{herre2}
\end{figure}

The Elbow Curve and the Silhouette analysis have been the methods employed to choose an appropriate value for the number of clusters. The elbow method runs k-means clustering on the dataset for a range of values of k (2-25) and, for each one, the average distances to the centroid across all data points is computed. The cut-off point where the diminishing SSE is no longer worth the additional number of clusters, indicates the optimal number of clusters. A comprehensive analysis has been performed to find the optimal number of clusters and the encoding strategy (see Figure \ref{herre2}) for results in Company1.

Because of the visual criteria’s subjectivity inherent to the Elbow method, Silhouette coefficient has been used the analytical method to contrast the result. The higher the coefficient, the more compact are the clusters and easily differentiable from the rest. The interpretation of the silhouette coefficient adopted on this work has been the following \cite{b33}:
\begin{itemize}
\item Sc $_{>}$ 0,7: strong cluster structure
\item Sc  $_{>}$ 0,5: reasonable cluster structure
\end{itemize} 

\begin{figure}[!t]
\centerline{\includegraphics[width=0.5\columnwidth]{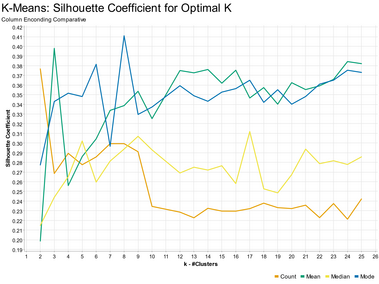}}
\caption{Comparison of Silhouette coefficient by encoding strategy (frequency, mean, median and mode). Company 1}
\label{herre4}
\end{figure}

Figure \ref{herre4} shows the silhouette coefficient computed for each number of clusters tested (parameter k) and the encoding strategy for Company1. In general, the Median, Mean and Mode strategies provides better results as their overall coefficient is higher than the Count strategy with any k value. This means that the clusters are more compact and easily differentiable from the rest.

According to the adopted criteria, at this stage of the work, the k-Means model chosen for Company1 has been the dataset encoded by Mode strategy and 8 clusters (SSE=208.459 Silhouette=0,41) and, for Company2, the dataset encoded by Median strategy and 5 clusters (SSE=267.784 Silhouette=0,34). In both cases, the coefficients obtained are below the threshold of a reasonable cluster structure.

Due to its high-demanding computational cost, as approximation to efficiently compute the coefficient in large datasets, a stratified random sampling (10\% of the total population) has been performed before its calculation \cite{b65}. This sampling method provides an unbiased and accurate representation of the a priori distribution of a feature in the underlying population. Given the poor values obtained (\textit{Silhouette} was below 0,5 for every number of clusters between 2 and 25 and for any of the encoding datasets), a coefficient’s full computation for the Company1 dataset (including univariate outliers) has shown quite similar results so the sampling method used for the \textit{Silhouette} computation shows as a good approximation. 

\subsubsection{k-Means – Optimised - Without univariate outliers}

\begin{figure}[!t]
\centerline{\includegraphics[width=0.5\columnwidth]{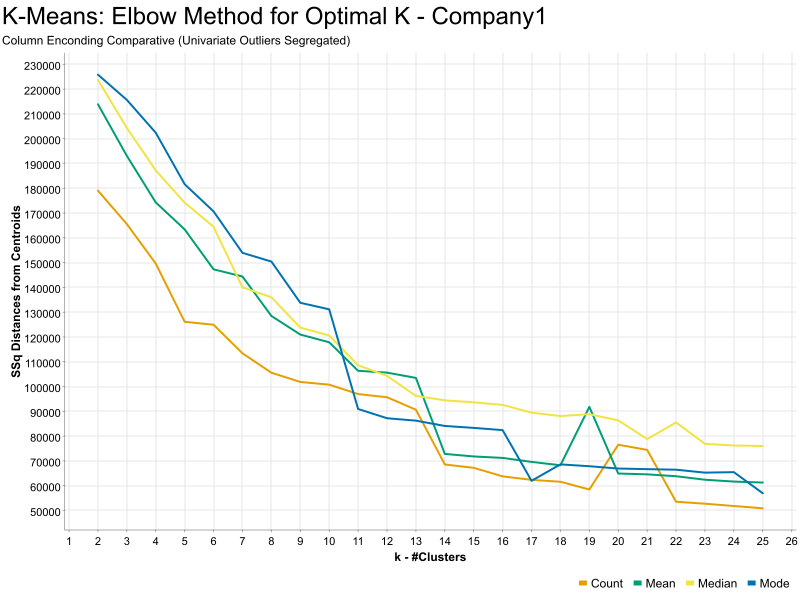}}
\caption{Company1 - Comparison of the Elbow Curves by encoding strategy (frequency, mean, median and mode) with univariate outliers segregated }
\label{herre6}
\end{figure}

\begin{figure}[!t]
\centerline{\includegraphics[width=0.5\columnwidth]{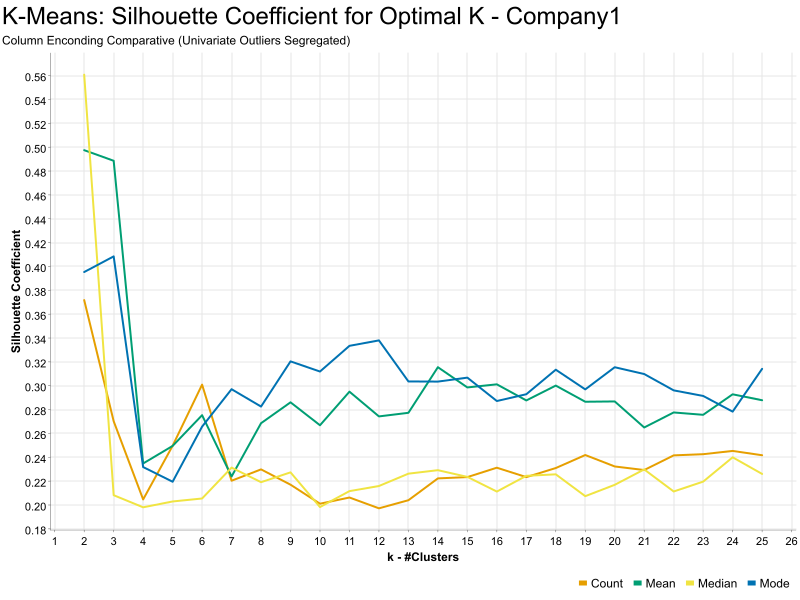}}
\caption{Company1 - Comparison of the Overall Silhouette coefficient – 10\% sampled by encoding strategy (frequency, mean, median and mode) with univariate outliers segregated }
\label{herre7}
\end{figure}

As seen in the previous step, the indicators of the cluster validity were not suitable according to commonly adopted criterion (reasonable cluster structure: Sc $_{>}$ 0.5). For this reason, the univariate outliers have been segregated from the datasets and the process has been repeated (previously, a re-encoding of the categorical features has been computed).
Figures \ref{herre6} and \ref{herre7} show the Elbow curves and the Overall Silhouette coefficient in relation to the number of clusters and for each encoding strategy for the Company1 dataset with the univariate outliers segregated. The Elbow curve for Count remains approximately at the same SSE levels while Mean, Median and Mode curves have decreased their sum of squared distances to centroids closer to Count values. As expected, filtering out the univariate outliers has a direct impact on the encoding strategies depending on the central trending of the features/columns.

The \textit{Silhouette} coefficient almost reaches an acceptable level in k=2-3 for the Mean and Median datasets although, in comparison with the results with no segregation of outliers (Figure \ref{herre7}), all the datasets decrease dramatically from this value of k. In this case, a low number of clusters is not adequate for our problem, given that they will contain a high number of transactions (e.g. Median dataset with k=2 clusters, the smaller cluster with 2217 transactions) whose review by an expert is unfeasible. Similar results have been obtained for Company2.

Filtering out the univariate outliers from the datasets should improve the quality indicators as long-distant transactions are no longer considered for the computation. However, in terms of clustering quality indicators, a significant improvement has not been achieved and the hyperparameters selected to model each company are those adopted without segregation of the univariate outliers).

\subsubsection{Isolation Forest}
\label{sec:IsoForest}

The KNIME nodes used are implemented using H2O \cite{b67}:
\begin{itemize}
\item Node H2O Isolation Forest Learner. The model is learned unsupervised from the input. The main training hyper-parameters used are the number of trees to build (nTree=100) and the subsampling size to train each tree (sample\_size=256); both are the defaults of the original paper \cite{b41}. A static random seed has been used.
\item Node H2O Isolation Forest Predictor.  The node applies the model resulting from the learner to an input dataset to predict anomalies or outliers. The output of the node will consist of:
\begin{itemize}
\item Prediction expressed as a normalized anomaly score. The higher the score,
the more likely it is an anomaly.\\
$Prediction =\frac{MaxPathLenth - MeanPathLength}{MaxPathLength - MinPathLength}$
\item Mean length of the predicted decision tree paths of each observation. The
shorter, the more likely it is an anomaly.\\
$MeanLength = \frac{PathLength}{nTree}$\\
\end{itemize}
\end{itemize}

Given that the dataset is not labelled and, thus, unsupervised algorithms are required, the learning and predictor stage of the algorithm has been applied upon each one of the whole companies’ datasets.

Due to the unknown proportion of outliers in the unlabelled datasets used, iForest cannot accurately set a threshold to determine whether a certain transaction is put into the anomaly candidates set. In our case, the anomaly threshold for the IForest algorithm is established as the minimum score of the 1\% of the transactions ordered descending by the prediction score (with a maximum of 500 transactions). This approach is based on the limited availability of the business specialists, and it can be revisited according to the actual effort they can dedicate. This threshold is computed as the 99th quantile of the Prediction provided by IForest and flagging true/false according to each transaction .

\subsection{Anomalies Prioritisation - Ensemble Algorithm}

As a preliminary step to establish a global anomalies prioritisation, the dataset with an optimal k-Means model has been chosen and, upon these, the adopted criteria to prioritise the anomalies have been the following:

\begin{enumerate}
\item[i] Low populated clusters. Initially, the clusters populated less than 1\% of the total dataset have been defined as the analytical threshold. The transactions belonging to those clusters have been flagged as “k-Means anomaly”.
\item[ii] Single transactions which differ from others in the same cluster. In other words, outliers within the clusters. The transactions with an individual Silhouette coefficient negative have been flagged as “Silhouette anomaly”.
\item[iii] Isolation Forest prediction. According to the effort-based approach, already explained in section \ref{sec:IsoForest}, those transactions beyond the 99th quantile of the algorithm prediction have been flagged as “iForest anomaly”. An absolute maximum of 500 transactions is established.
\item[iv] Univariate anomalies. In the case of not having segregated the univariate outliers from the dataset, those transactions with at least one univariate anomaly -according to z-Score or DBSCAN - have been flagged as “univariate outlier anomaly”.
\end{enumerate}

Criteria i) and ii) correspond to the 2\textsuperscript{nd} and 3\textsuperscript{rd} categories of clustering-based techniques for anomaly detection proposed by \cite{b24}. Some studies have adopted this categorisation to establish the key assumptions to identify anomalies \cite{b20,b68}.

In a second step, a re-prioritisation can be carried out in case of an excessive number of anomalies -with respect to the specialist’s availability to review the results-. For example, if the number of “k-Means anomalies” must be reduced, they can be ordered by the individual Silhouette coefficient (the lower, the more critical) or by the iForest prediction (the higher, the more critical). 

A simple global prioritisation, a kind of independent ensemble, has been implemented flagging the transactions with the number of anomaly groups they belong to and ordering them (descending).  The resulting prioritised groups are shown in Table \ref{tab:tab5} and Table \ref{tab:tab6}. For a better understanding of the anomaly location, the distribution of the prioritised groups within the k-Means clusters is provided.

\begin{table}[]
\centering
\caption{Company1 - Mode dataset. k=8. Global prioritisation groups by type of anomaly}
%\begin{tabular}{|c|c|c|c|c|c|c|c|c|c|c|c|c|c|}
\setlength{\tabcolsep}{3pt}
\begin{tabular}{lllllllllllllll}
\hline
\rotatebox[origin=c]{90}{Anomaly Priority} &
\rotatebox[origin=c]{90}{k-Means Anomaly}   &
  \rotatebox[origin=c]{90}{Silhouette Anomaly} &
  \rotatebox[origin=c]{90}{iForest Anomaly} &
  \rotatebox[origin=c]{90}{Univariate Anomaly} &
  \rotatebox[origin=c]{90}{cluster 0} &
  \rotatebox[origin=c]{90}{cluster 1} &
  \rotatebox[origin=c]{90}{cluster 2} &
  \rotatebox[origin=c]{90}{cluster 3} &
  \rotatebox[origin=c]{90}{cluster 4} &
  \rotatebox[origin=c]{90}{cluster 5} &
  \rotatebox[origin=c]{90}{cluster 6} &
  \rotatebox[origin=c]{90}{cluster 7} &
  \rotatebox[origin=c]{90}{Total} \\ \hline
\cellcolor[HTML]{F8A102}2 & No & No  & Yes & Yes & 1    & 19   & 3   & 4    & 14   & 0    & 0    & 3    & 44    \\ \hline
\cellcolor[HTML]{F8A102}2 & No & Yes & No  & Yes & 34   & 0    & 3   & 11   & 0    & 0    & 3    & 100  & 151   \\ \hline
\cellcolor[HTML]{F8A102}2 & No & Yes & Yes & No  & 0    & 0    & 0   & 0    & 0    & 0    & 0    & 7    & 7     \\ \hline
\cellcolor[HTML]{32CB00}1 & No & No  & No  & Yes & 1411 & 1484 & 305 & 335  & 290  & 226  & 202  & 273  & 4526  \\ \hline
\cellcolor[HTML]{32CB00}1 & No & No  & Yes & No  & 0    & 0    & 0   & 14   & 158  & 6    & 9    & 45   & 232   \\ \hline
\cellcolor[HTML]{32CB00}1 & No & Yes & No  & No  & 0    & 0    & 0   & 89   & 0    & 0    & 9    & 768  & 866   \\ \hline
\cellcolor[HTML]{036400}0 & No & No  & No  & No  & 0    & 0    & 0   & 5150 & 5263 & 4352 & 4183 & 2947 & 21895 \\ \hline
\end{tabular}%

\label{tab:tab5}
\end{table}

\begin{table}[]
\centering
\caption{Company2 - Median dataset. k=5. Global prioritisation groups by type of anomaly}
%\begin{tabular}{|c|c|c|c|c|c|c|c|c|c|c|}
\setlength{\tabcolsep}{3pt}
\begin{tabular}{lllllllllll}
\hline
\rotatebox[origin=c]{90}{Anomaly Priority} &
\rotatebox[origin=c]{90}{k-Means Anomaly}   &
  \rotatebox[origin=c]{90}{Silhouette Anomaly} &
  \rotatebox[origin=c]{90}{iForest Anomaly} &
  \rotatebox[origin=c]{90}{Univariate Anomaly} &
  \rotatebox[origin=c]{90}{cluster 0} &
  \rotatebox[origin=c]{90}{cluster 1} &
  \rotatebox[origin=c]{90}{cluster 2} &
  \rotatebox[origin=c]{90}{cluster 3} &
  \rotatebox[origin=c]{90}{cluster 4} &
  \rotatebox[origin=c]{90}{Total} \\ \hline
\cellcolor[HTML]{FE0000}3 & Yes & Yes & No  & Yes & 60 & 0    & 0    & 0    & 0    & 60    \\ \hline
\cellcolor[HTML]{F8A102}2 & No  & No  & Yes & Yes & 0  & 10   & 0    & 21   & 7    & 38    \\ \hline
\cellcolor[HTML]{F8A102}2 & No  & Yes & No  & Yes & 0  & 4    & 0    & 0    & 71   & 75    \\ \hline
\cellcolor[HTML]{F8A102}2 & Yes & No  & No  & Yes & 28 & 0    & 0    & 0    & 0    & 28    \\ \hline
\cellcolor[HTML]{32CB00}1 & No  & No  & No  & Yes & 0  & 919  & 633  & 796  & 882  & 3230  \\ \hline
\cellcolor[HTML]{32CB00}1 & No  & No  & Yes & No  & 0  & 147  & 10   & 173  & 11   & 341   \\ \hline
\cellcolor[HTML]{32CB00}1 & No  & Yes & No  & No  & 0  & 0    & 0    & 0    & 355  & 355   \\ \hline
\cellcolor[HTML]{036400}0 & No  & No  & No  & No  & 0  & 8990 & 8437 & 8456 & 7928 & 33811 \\ \hline
\end{tabular}%

\label{tab:tab6}
\end{table}

Other approaches are possible to set analytical rules and merge the individual algorithms’ results. \cite{b40} uses these kind of approaches to combine anomalies detected in audit logs for application systems by LOF and DBSCAN algorithms. 

\subsection{Model Interpretability}

Interpretable Machine Learning (IML) methods can be used to discover knowledge, debug or justify the model and its predictions, and control and improve the model \cite{b47}. 
Methods that study the sensitivity of an ML model are mostly model-agnostic and work by manipulating input data and analysing the respective model predictions. These IML methods often treat the ML model as a closed system that receives feature values as an input and produces a prediction as output. 

\begin{figure*}[!t]
\centerline{\includegraphics[width=\textwidth]{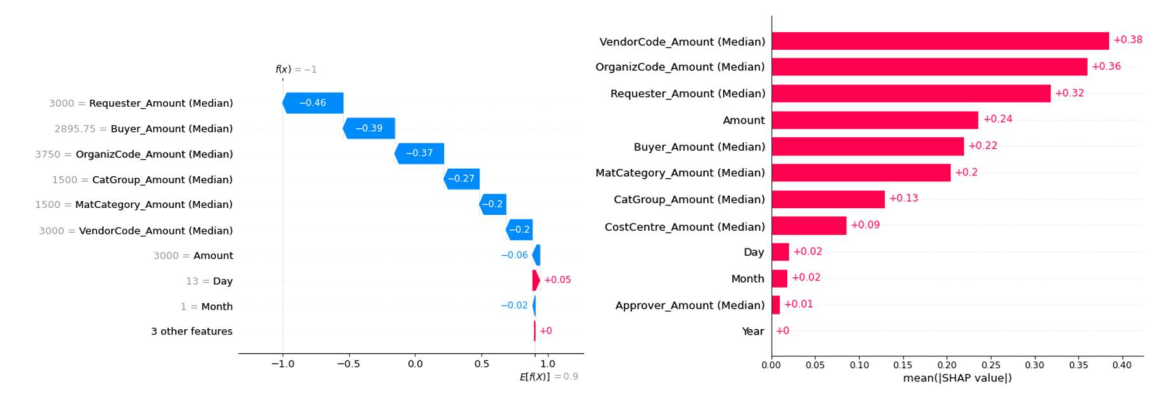}}
\caption{Company2. Example of SHAP values. (left) Force plot for individual explicability. (right) Bar plot for feature explicability }
\label{herre8}
\end{figure*}

Some of the available explicability tools (Shapley values, SHAP and Lime, \cite{b47}) has been tested to provide an overview of how they can help the business specialists in the task of understanding why an outlier has been flagged \cite{b47}. A further investigation needs to be accomplished to define the better and most reliable explicability method to support the specialists in their analysis. Figure \ref{herre8} shows a test using the SHAP package in Python. In the sklearn implementation of iForest, the lower the score is, the more abnormal \cite{b42}. For the example shown in the figure, the algorithm has been applied to the Company2 dataset, disclosing that the requester, the buyer and the department (organization code) have the major contributions to the anomaly deviation with respect to the dataset average prediction. 

\section{Conclussion}

In this paper, Machine Learning based anomaly detection has been performed on real purchase datasets, in order to provide a disciminant tool for fraud detection in auditory processes. A combination of several anomaly detection techniques (ADT) is proposed: univariate detection (through z-score and DBSCAN), low populated clusters and the negative Silhouette coefficients defined by k-Means, and the Isolation Forest.

The most suitable models generated by the k-Means algorithm presented a low clustering quality according to the overall Silhouette index (below 0,5). Complementary analysis, as using datasets without univariate outliers or performing a full computation of the Silhouette coefficient, has not improved the quality measures. 

Concerning the adopted assumptions to identify anomalies, low populated clusters were assumed as collective outliers, flagging transactions belonging to clusters representing less than 1\% of the total dataset. To provide further information and to better assess the results, the rows (transactions) with negative Silhouette index were flagged as potential contextual anomalies. To complement the cases with a different approach and enrich the model with a wide variety of the detectable anomalies’ nature, the transactions with the highest prediction computed through Isolation Forest algorithm were selected (those higher than 99th quantile). Finally, the preliminary univariate outlier identification performed during the EDA phase was retained in the output, flagging transaction with, at least, one univariate outlier on its features.

As showed in this work, the use of the KNIME has been revealed as an easy-to-use and well-documented tool. It enables the use of a wide range of Machine Learning techniques without much effort, and makes possible to automatize processes, being a key aspect in iterative investigations. Python integration has allowed to verify some the results and improve the output’s look and feel.

The work has revealed the extensive possibilities that Machine Learning, and particularly clustering, offers to unsupervised learning problems as the procurement process. Among others, some axes of future investigation are testing of other encoding options for categorical features, performing a time analysis for the purpose of detecting stational behaviours (splitting purchase transactions in a short period of time is among the major internal control concerns in a company), extending the use of clustering algorithms to others (such as DBSCAN, Fuzzy c-Means - overlapping clustering -, hierarchical clustering or Self-Organised Maps -SOM-). This certainly will extend the nature of the anomalies addressed by the proposed methodology.

%\appendices

\section*{Acknowledgment}

    This work is partially supported by MCIN/AEI/ 10.13039/501100011033 by “ERDF A way of making Europe”, grant PID2021-127946OB-I00. The author, S. Pérez-Díaz, is partially supported by Ministerio de Ciencia, Innovación y Universidades - Agencia Estatal de Investigación/PID2020-113192GB-I00 (Mathematical Visualization: Foundations, Algorithms and Applications).

\end{document}